\documentclass[pra,aps,reprint,a4paper,superscriptaddress,longbibliography,floatfix]{revtex4-1}

\usepackage{bm}
\usepackage{times}
\usepackage{epsfig}
\usepackage{amsfonts}
\usepackage{amsmath}
\usepackage{amssymb}
\usepackage{amsthm}
\usepackage{color}
\usepackage{multirow}
\usepackage[colorlinks=true,linkcolor=blue,citecolor=magenta,urlcolor=blue]{hyperref}

\begin{document}


\title{Bell nonlocality with intensity information only}


\author{Ari Patrick}
\email{aripatrick@usp.br}
\affiliation{Departamento de F\'{\i}sica-Matem\'atica, Instituto de F\'{\i}sica, Universidade de S\~ao Paulo, S\~ao Paulo 05508-090, S\~ao Paulo, Brazil}

\author{Ad\'an Cabello}
\email{adan@us.es}
\affiliation{Departamento de F\'{\i}sica Aplicada II, Universidad de
	Sevilla, E-41012 Sevilla, Spain}
\affiliation{Instituto Carlos~I de F\'{\i}sica Te\'orica y Computacional, Universidad de
	Sevilla, E-41012 Sevilla, Spain}


\begin{abstract}
We address the problem of detecting bipartite Bell nonlocality whenever the only experimental information are the intensities produced in each run of the experiment by an unknown number of particles. We point out that this scenario naturally occurs in Bell experiments with parametric down-conversion when the crystal is pumped by strong pulses, in Bell tests with distant sources and in which particles suffer different delays during their flight, in Bell experiments using living cells as photo detectors, and in Bell experiments where the pairing information is physically removed. We show that, although Bell nonlocality decreases as the number of particles increases, if the parties can distinguish arbitrarily small differences of intensities and the visibility is larger than $0.98$, then Bell nonlocality can still be experimentally detected with fluxes of up to $15$~particles. We show that this prediction can be tested with current equipment in a Bell experiment where pairing information is physically removed, but requires the assumption of fair sampling. 
\end{abstract}


\maketitle


\section{Introduction}


\subsection{Motivation}
\label{mot}


Bell nonlocality, that is, the violation of Bell inequalities \cite{Bell64}, is one of the most characteristic signatures of quantum theory and has a wide range of applications for communication and computation \cite{BCPSW14}. However, Bell nonlocality vanishes if the following hold true. 

(i) The only experimental information available to the parties (Alice and Bob) are the intensities registered by their respective detectors.

(ii) Intensities are produced by continuous fields (rather than by discrete particles).

(iii) The parties can only measure changes in intensity values of the order of $\sqrt{N}$, where $N$ is the number of particles.

To explain why, let us consider the simplest Bell scenario: two parties, each of them with two measurement options, $x\in\{1,2\}$ for Alice and $y\in\{1,2\}$ for Bob, and each measurements with two possible outcomes, $a\in\{0,1\}$ for Alice's measurements and $b\in\{0,1\}$ for Bob's. Alice and Bob, using classical communication, can compute the marginal probability densities $p(I_{a|x},I_{b|y})dI_{a|x} dI_{b|y}$, where $I_{a|x}$ and $I_{b|y}$ are the intensities registered by, respectively, Alice's and Bob's detectors. If conditions (i) to (iii) hold, then, in the limit $N \gg 1$, consistency with classical physics forces this set of marginal distributions to admit a local hidden variable model for the intensities. In this case, it is said that the intensities exhibit ``macroscopic locality'' \cite{NW09}.


\begin{figure}[t]
	\includegraphics[width=8.2cm]{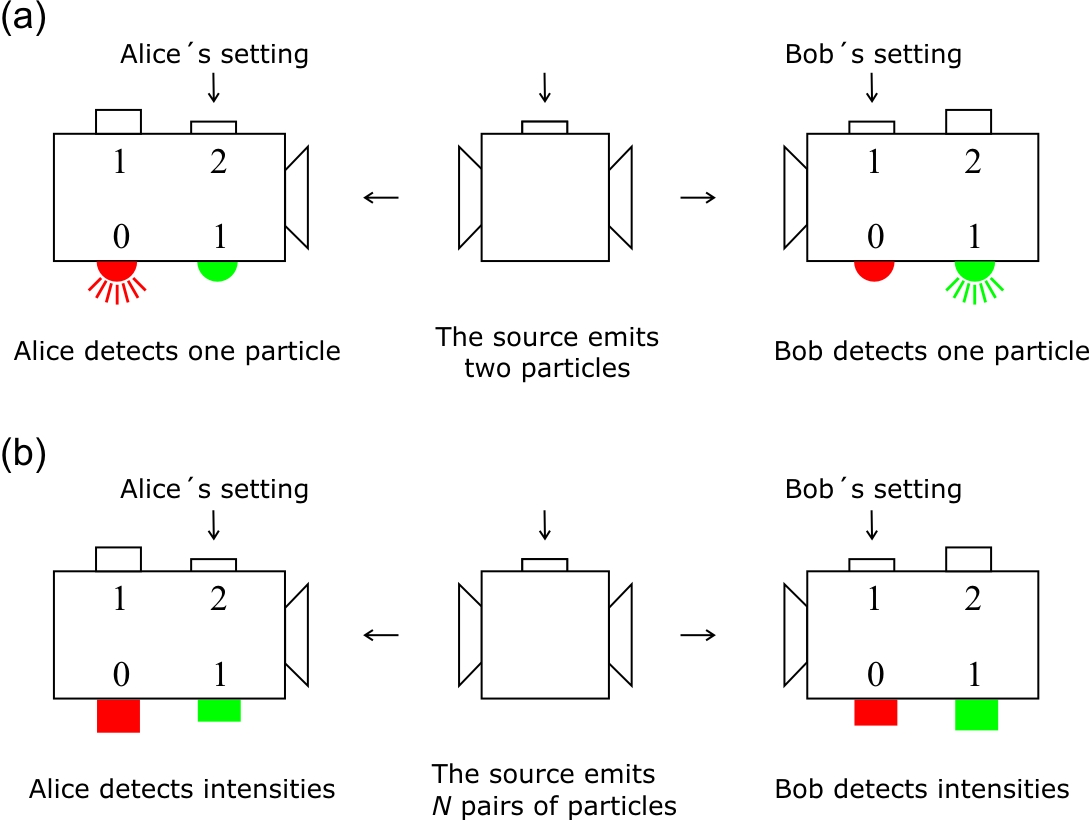}
	\caption{(a) Standard Bell test in which the source emits a pair of entangled particles and each of Alice and Bob detects one of them after choosing their respective measurement setting. (b) Bell test studied here, in which the source emits $N$ pairs of entangled particles and each of Alice and Bob only detect the intensities produced by a flux of $N$~particles after choosing their respective measurement setting.}
	\label{Fig1}
\end{figure}


Here, we investigate what happens when condition~(i) holds but conditions~(ii) and (iii) do not. While in a standard Bell test, every time one party chooses a measurement setting detects one particle, as illustrated in Fig,~\ref{Fig1}(a), in this work we assume that, instead of that, each party detects $N>1$ particles. In addition, we assume that each party only has access to the intensities these $N$ particles have produced and may not even know in advance the value of $N$, which may also be slightly different in different runs of the experiment. See Fig.~\ref{Fig1}(b). However, we assume that the parties can distinguish any small difference of intensity between their two detectors. The problem we address is what happens with the violation of a Bell inequality when, in each run of a Bell experiment (i.e., when Alice has chosen to measure $x$ and Bob has chosen to measure $y$), both detectors on each party might detect some photons, and they have to provide an outcome based on the detection. 

Our motivation for investigating this problem is twofold. From a fundamental perspective, we know that the electromagnetic field behaves as made of individual packets called photons and, therefore, intensities can be seen as produced by an accumulation of discrete particles. That is, we know that, from a fundamental perspective, condition~(ii) does not hold.

In addition, from a practical perspective, there are several scenarios in which the only experimental information available to Alice and Bob are intensities while Alice and Bob can still distinguish small differences of intensities. We have identified the following examples.

(1) Bell experiments, where the source uses a parametric down-conversion process \cite{Kwiat} and the nonlinear crystal is pumped with strong pulses, so each pulse sometimes produces $N$~pairs of entangled photons rather than a single pair. Here, we assume that the detectors can collect all these photons. This is possible, using, for instance, special arrays of nanowire detectors \cite{Wamsley}. Notice that in these experiments the visibility of the state is intrinsically affected by the thermal nature of the source thus it may be difficult to achieve very high visibilities.

(2) Bell experiments where the source of entangled pairs is moving and far from the detectors. In addition, there may be disturbances during the propagation of the particles which make it impossible to identify which particle of Alice is entangled with which particle of Bob. For example, this happens when the source randomly oscillates in the direction of propagation of the particles at higher speeds than the speed of propagation of the particles and/or when particles propagate at different speeds due to local disturbances. Recall that the propagating particles may not be photons and also that photons can slow down when passing through a slow-light medium \cite{Hau11}.
A similar situation is that of future satellite-to-ground Bell tests in which the source is in the satellite and both Alice and Bob are in the ground (in current satellite-to-ground Bell tests, Alice is in the satellite with the source \cite{Pan1}). Other example would be Bell tests with hypothetical cosmic sources of entanglement.

(3) In Bell experiments that use hybrid photo detectors that incorporate living cells. For example, rod photoreceptor cells taken from the eye of a frog \cite{K12}. There, each rod has an outer segment that contains rhodosin molecules that undergoes a chemical change when exposed to light. This results in an electrical signal that is picked up by the nervous system and relayed to the brain. When submitted to a flux of photons, each photon interacts with just one rhodosin molecule \cite{K12}. If one can distinguish which electrical signal (the one corresponding to $0$ or the one corresponding to $1$) corresponds to a higher intensity, then we are in the case that we are considering. If this distinction would be possible in the brain, we could detect Bell nonlocality using human eyes and without needing entanglement amplification (as in \cite{SBBGS09,PSSZG11}).

(4) In Bell experiments where pairing information is {\em physically} erased. We introduce and discuss one of these experiments in Sec.~\ref{exp}.


\subsection{Relation with previous works}
\label{prev}


Bell experiments with only intensity information have been discussed before for different purposes \cite{NW09,BBBGPS08,PCBCSSK17,ZCBGS17}. Here, we point out some differences between these works and the present work. 

Bancal {\em et al.} \cite{BBBGPS08} defined a ``multipair [bipartite] scenario'' as one where there are $N$ independent sources and each produces a pair of particles (or, equivalently, there is a single source producing $N$ pairs), the pairing between Alice's and Bob's particles is lost during their transmission, and each party measures all their incoming particles in the same basis. For these scenarios, they showed that several strategies allow for Bell nonlocality. However, none of the scenarios 1 to 4 above, is mentioned. 

In a follow-up paper, Poh {\em et al.} \cite{PCBCSSK17} showed a strategy that leads to a larger violation for multipair scenarios and include an experiment to demonstrate this advantage. These works assumed that $N$ is constant during the experiment and known by the parties, so adaptive strategies depending on $N$ are possible. 

In contrast to that, in the physical Bell tests with only intensity information we are interested in (scenarios 1 to 4), $N$ can be unknown to the parties and can slightly change in each run of the experiment. Hence, the parties may have to chose their measurements not knowing $N$. This is the reason why in this paper we fix the local measurements and do not consider strategies that depend on $N$. On the other hand, while the experiment in \cite{PCBCSSK17} is a Bell experiment {\em with particle pairing information} but postprocessed to simulate the loss of pairing information, in this work we emphasize that there are actual Bell experiments in which the pairing information is {\em physically} removed either effectively, as in examples 1 to 3, or fundamentally, as in example 4 (see Sec.~\ref{exp}).

To conclude, in \cite{ZCBGS17} Zhou {\em et al.} also considered Bell experiments with only detection intensities. However, their aim was to discuss a version of the macroscopic locality ``principle'' introduced in \cite{NW09} for trying to single out Bell nonlocal correlations from fundamental principles. For a recent development on this subject, see \cite{Cabello19}.


\subsection{Bell vs contextuality experiments with intensities}
\label{cont}


Before continuing, it is important to stress that the limitation that the only experimental information available is the intensities in each detector creates a fundamentally different problem in Bell experiments than in Kochen-Specker contextuality experiments \cite{Cabello08}. In the second case, space-like separation plays no role so in the Kochen-Specker contextuality experiment equivalent to the simplest Bell inequality experiment Bob's measurement can be time-like separated from Alice's. This allows us to encode each of Alice's discrete outcomes in an extra degree of freedom (for example, path \cite{ARBC09}, time \cite{AACB13}, or polarization \cite{MANCB14}) before Bob's measurement is performed. This allows us to guide the flux of discrete particles to four different detectors: the first, corresponding to the case the outcomes of Alice and Bob are $0,0$ (respectively); the second for the case $0,1$; the third for $1,0$; and the fourth detector for $1,1$. This trick of encoding Alice's outcomes in an extra degree of freedom before performing Bob's measurement allows us to recover quantum contextual correlations even using classical microwaves and classical light \cite{FBV16,ZXX19}. However, this trick is not possible when there is space-like separation between Alice's and Bob's measurements as is the case in Bell experiments.


\subsection{Structure of the paper}


The structure of the paper is as follows. In Sec.~\ref{ic}, we begin by assuming that both the state preparation and the detection efficiency are perfect. That is, that there is no noise and each of Alice and Bob detects $N$ particles. Then, in Sec.~\ref{nd}, we study the effect of noise, and in Sec.~\ref{if}, we study the effect of imperfect detection. In Sec.~\ref{exp}, we propose testing this prediction in an experiment in which pairing information is physically erased. 


\section{Results}


\subsection{Ideal case}
\label{ic}


We consider the simplest Bell inequality, the Clauser-Horne-Shimony-Holt \cite{CHSH69} Bell inequality, in the version proposed by Zohren and Gill in \cite{ZG08}, namely,
\begin{equation}
S \ge 1,
\end{equation}
with
\begin{eqnarray}\label{eq_1}
S&=&p(01|22)+p(10|12)+p(01|11) \nonumber \\
&&+p(11|21)+p(10|21)+p(00|21),
\end{eqnarray}
where $p(ab|xy)$ is the probability of obtaining outcomes $a$ and $b$ for measurements $x$ and $y$, respectively. 

In the case of $N=1$, the maximum quantum violation is 
\begin{equation}
S=\frac{3-\sqrt{2}}{2} \approx 0.793,
\end{equation} 
and is achieved, for example, with the state
\begin{equation}
\label{state}
|\phi^+\rangle=\frac{1}{\sqrt{2}}(|0\rangle_A \otimes |0\rangle_B+|1\rangle_A \otimes |1\rangle_B),
\end{equation}
where, e.g., $|0\rangle_A$ denotes that Alice's particle is in the state represented by the vector $\left( \begin{array}{c} 1 \\ 0
\end{array}\right)$ and $|1\rangle_B$ denotes that Bob's particle is in the state represented by $\left( \begin{array}{c} 0 \\ 1
\end{array}\right)$,
and the following measurement settings:
\begin{subequations}
	\label{settings}
	\begin{align}
	&M_{x=1} = \frac{1}{2}(\openone - \sigma_x), \;\;\;\;\;\;\;\;\;\;\;\;\;\;M_{x=2} = \frac{1}{2}(\openone - \sigma_y),\\
	&M_{y=1} = \frac{1}{2}\left(\openone - \frac{\sigma_x + \sigma_y}{\sqrt{2}}\right), \;\;\;M_{y=2} = \frac{1}{2}\left(\openone - \frac{\sigma_x - \sigma_y}{\sqrt{2}}\right),
	\end{align}
\end{subequations}
where $\openone$ denotes the identity matrix and $\sigma_n$ the Pauli matrix in the direction $n$. Each of these observables has two possible outcomes: $0$ and $1$, corresponding to the eigenvalues of the operator that represents the observable.

Now we consider the case in which, in each run of the Bell experiment, each of Alice and Bob receives a number $N>1$ of particles every time they chose their measurement. $N_0$ of the particles end up in the detector corresponding to the outcome $0$ of the measurement and $N_1=N-N_0$ of the particles end up in the detector corresponding to the outcome $1$. There is no information about the order in which the particles arrived. The only information are the intensities $I_0 = k N_0$ and $I_1 = k N_1$ in each detector. Using this information, each of the parties, without communicating with the other party, should provide an outcome $0$ or $1$. The question is which is the strategy that better preserves Bell nonlocality.

After checking all possible alternatives, we found that an optimal strategy is the one in which each party outputs the detector that has higher intensity. That is, if $I_0 > I_1$, then the party outputs $0$, while, if $I_0 \le I_1$, then the party outputs $1$.

Then, for example, to compute $p_N(ab|xy)$, defined as the probability of Alice yielding the outcome $a$ for measurement $x$, and Bob yielding the outcome $b$ for measurement $y$, when each of them detect $N$~particles, we have to sum the probabilities of all the possible ways in which $N$ particles in Alice's side and $N$ particles in Bob's side may have induced Alice to output $a$ and Bob to output $b$. 

For example, for $N=2$,
\begin{eqnarray}\label{eq_10}
p_2(01|xy) &=& p(00|xy)p(01|xy) + p(01|xy)p(00|xy) \nonumber\\
&+& p(01|xy)p(01|xy),
\end{eqnarray}
where $p(00|xy)p(01|xy)$ is the probability that the first pair of particles ended in detector $0$ for Alice and Bob, while the second pair ended in Alice's detector $0$ and Bob's detector $1$.

For arbitrary $N$,
	\begin{subequations}
		\begin{align}
	&p_N(00|xy) = \sum_{\sum_{i=1}^{N} a_i < \frac{N}{2}} \; \sum_{\sum_{j=1}^{N} b_j < \frac{N}{2}} \prod_{k=1}^{N}p(a_kb_k|xy),\\
	&p_N(01|xy) = \sum_{\sum_{i=1}^{N} a_i < \frac{N}{2}} \; \sum_{\sum_{j=1}^{N} b_j \ge \frac{N}{2}} \prod_{k=1}^{N}p(a_k b_k|xy),\\
	&p_N(10|xy) = 
	\sum_{\sum_{i=1}^{N} a_i \ge \frac{N}{2}}\;\sum_{\sum_{j=1}^{N} b_j < \frac{N}{2}} \prod_{k=1}^{N}p(a_k b_k|xy),\\
	&p_N(11|xy) = \sum_{\sum_{i=1}^{N} a_i \ge \frac{N}{2}} \; \sum_{\sum_{j=1}^{N} b_j \ge \frac{N}{2}} \prod_{k=1}^{N}p(a_k b_k|x y).
		\end{align}
	\end{subequations}

\begin{figure}[tb]
	\centering
	\includegraphics[width=8.2cm]{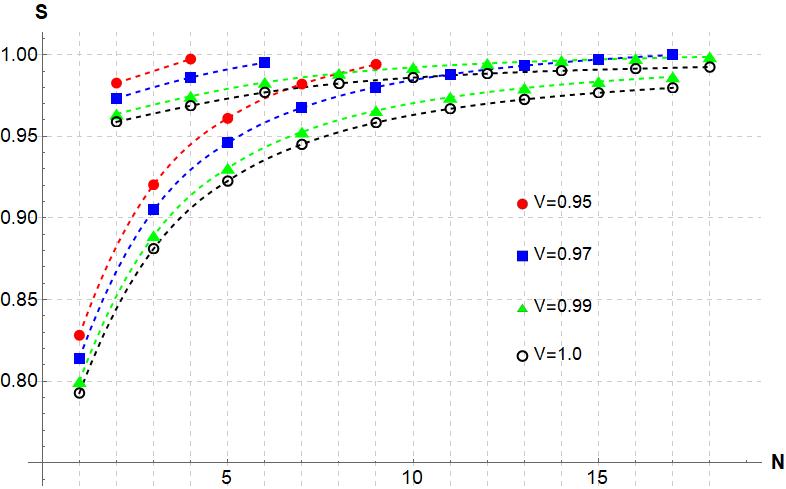}
	\caption{Maximum quantum violation of the Bell inequality $S_N \ge 1$ as a function of the number $N$ of particles detected by each party for states of the form~(\ref{eq_rho}), for different values of $V$. For a given $V$, the violation is larger when $N$ is odd. For $V=0.95$, the violation vanishes for $N>9$, if $N$ is odd, and for $N>4$ if $N$ is even.
	For $V=0.97$, the violation vanishes for $N>17$, if $N$ is odd, and for $N>6$ if $N$ is even. For $V\ge 0.99$, there is always violation (although very small) at least up to $N=18$. Obtaining the maximum quantum violation for higher values of $N$ requires computing power that exceeds our capabilities.}
	\label{Fig2}
\end{figure}

Therefore, we can define
\begin{eqnarray}\label{eq_N}
S_N&=&p_N(01|22)+p_N(10|12)+p_N(01|11) \nonumber \\
&&+p_N(11|21)+p_N(10|21)+p_N(00|21)
\end{eqnarray}
and compute the maximum quantum violation of the Bell inequality $S_N \ge 1$ as a function of $N$. The result of our calculations is presented in Fig.~\ref{Fig2} (the case $V=1.0$).


Notice that the maximum quantum violation depends on whether $N$ is even or odd. The violation of the Bell inequality is larger for $N$~odd. This is due to the fact that only when $N$ is odd are the intensities in both detectors always unequal, so the strategy of yielding the largest intensity as outcome partially keeps the quantum behavior. In contrast, when $N$ is even, the intensities in both detectors are sometimes equal and, then, yielding~$1$ as the outcome destroys any quantum correlation and degrades the violation.


\subsection{The effect of noise}
\label{nd}


So far, we assumed that the state is perfect. Here, we examine the case in which the state is affected by some amount of white noise. Specifically, we assume that the prepared state is 
\begin{equation}\label{eq_rho}
\rho = V |\phi^+\rangle\langle \phi^+|+ (1-V) \frac{\openone}{4},
\end{equation}
where $V$, sometimes referred to as the visibility of the state, is not $1$. For state-of-the-art photonic experiments $V \ge 0.98$ \cite{PJCC15}. Here, we computed the maximum quantum violation for $V=0.95$, $V=0.97$, and $V=0.99$. The results are presented in Fig.~\ref{Fig2} (red circles, blue squares, and green triangles, respectively).

The maximum quantum violation depends on whether $N$ is even or odd. Higher noise makes the violation disappear for large values of~$N$. Interestingly, our results suggest that, with $V \approx 0.99$, it is possible to experimentally observe a statistically significant violation of the Bell inequality even with fluxes of up to $15$~particles.


\subsection{Inefficient detectors}
\label{if}


So far, we assumed that the detectors capture the $N$~pairs of particles emitted by the source. However, in actual experiments detectors only capture a fraction of the particles. Therefore, an interesting question is how robust the violation of inequality $S_N \ge 1$ is when some of the particles are missing but the parties keep following the same strategy. That is, if one party observes $I_0 > I_1$ in its detectors, then the party outputs~$0$, and, if it observes $I_0 \le I_1$, then the party outputs~$1$.

Here, we obtain the minimum detection efficiency $\eta_{\text{min}}$ needed to observe the violation of inequality $S_N \ge 1$ using the strategy mentioned above, as a function of $N$. The detection efficiency is the ratio between the number of particles detected by a detector and the number of particles emitted towards that detector. We assume that the source is heralded, $V=1$, all detectors have the same detection efficiency $\eta$, and there are no dark counts during the experiment.

For $N=1$, the strategy described above is equivalent to yielding outcome $1$ when no detection occurs. Then, we have the following cases. 

1.1 With probability $\eta^2$, both parties detect its particle. For this subensemble, $S_1=\frac{3-\sqrt{2}}{2}$. 

1.2 With probability $\eta (1-\eta)$, Alice detects and Bob does not. Therefore, Bob observes $I_0=I_1$ and always outputs~$1$. Therefore, for this subensemble, $S_1=p_A(0|2)+0+p_A(0|1)+p_A(1|2)+0+0=\frac{1}{2}+0+\frac{1}{2}+\frac{1}{2}+0+0=\frac{3}{2}$, where $p_A(0|2)$ is the probability that Alice finds the particle in detector~$0$ (and then outputs~$0$) when she measures~$2$. 

1.3 With probability $(1-\eta) \eta$, Alice does not detect and Bob detects. Therefore, Alice observes $I_0=I_1$ and always outputs~$1$. Therefore, for this subensemble, $S_1=0+p_B(0|2)+0+p_B(1|1)+p_B(0|1)+0=\frac{3}{2}$. 

1.4 Finally, with probability $(1-\eta)^2$, neither Alice nor Bob detects, so each of them always outputs $1$. For this subensemble, $S_1=0+0+0+1+0+0=1$. 

Therefore, $\eta_{\text{min}}$ follows from demanding that
\begin{equation}
\eta^2 \left( \frac{3-\sqrt{2}}{2} \right) + 2 \eta (1-\eta) \frac{3}{2} + (1-\eta)^2 < 1,
\end{equation}
which implies
\begin{equation}
\eta_{\text{min}} (N=1) = \frac{2}{1+\sqrt{2}} \approx 0.828.
\end{equation}
That is, there is Bell nonlocality (without making the fair sampling assumption; see below) if the detection efficiency is higher than this value. This value coincides with the one obtained for $N=1$ after optimizing over all strategies \cite{GM87}.




Let us now suppose that $N=2$. Then, we have the following cases.

2.1 With probability $\eta^4$, each of Alice and Bob detects the two particles. For this subensemble, $S_2=\frac{21}{16}-\frac{1}{2\sqrt{2}}$, where $S_2$ is defined in Eq.~(\ref{eq_N}).

2.2 With probability $2 \eta^3 (1-\eta)$, Alice detects the two particles and Bob only detects one (and thus he outputs the detector in which he found the particle). The factor of $2$ comes from the fact that the particle that Bob detects can be the one of the first part or the one of the second pair. To compute the value of $S_2$ for this subensemble, let us assume that Bob detects the first particle but not the second (the value of $S_2$ is the same if Bob detects the second but not the first one). Then,
\begin{eqnarray}
S_2&=&p(01|22)p_A(0|2) \nonumber \\
 &&+p(00|12)p_A(1|1)+p(10|12)p_A(0|1)+p(10|12)p_A(1|1) \nonumber \\
 &&+p(01|11)p_A(0|1) \nonumber \\
 &&+p(01|21)p_A(1|2)+p(11|21)p_A(0|2)+p(11|21)p_A(1|2) \nonumber \\
 &&+p(00|21)p_A(1|2)+p(10|21)p_A(0|2)+p(10|21)p_A(1|2) \nonumber \\
 &&+p(00|21)p_A(0|2) \nonumber \\
 &=& \left[p(01|22)+p(00|12)+p(10|12)+p(10|12) \right. \nonumber \\
 &&+p(01|11)+p(01|21)+p(11|21)+p(11|21) \nonumber \\
 &&\left. +p(00|21)+p(10|21)+p(10|21)+p(00|21) \right] \frac{1}{2} \nonumber \\
 &=&\frac{6-\sqrt{2}}{4}.
\end{eqnarray}
For example, $p(01|xy)p_A(0|x)$ is the probability that, for this subensemble, Alice outputs~$0$ and Bob outputs $1$ when they measure $x$ and $y$, respectively. This follows from the fact that Alice only outputs~$0$ when she finds the two particles in detector~$0$, and Bob only outputs~$1$ when he finds his (first) particle in detector~$1$ (the second particle is undetected). Similarly, $p(00|xy)p_A(1|x)+p(10|xy)p_A(0|x)+p(10|xy)p_A(1|x)$ is the probability that Alice outputs~$1$ and Bob outputs~$0$, since Alice outputs~$1$ when she finds one or the two particles in detector~$1$, and Bob only outputs~$0$ when he finds his particle in detector~$0$ (the second particle is undetected). 

2.3 With probability $2 \eta^3 (1-\eta)$, Alice detects one particle (and thus she outputs the detector in which she found the particle) and Bob detects the two particles. To compute the value of $S_2$ for this subensemble, let us assume that Alice detects the first particle but not the second (the value of $S_2$ is the same in the other case). Then,
\begin{eqnarray}
S_2&=&p(00|22)p_B(1|2)+p(01|22)p_B(0|2)+p(01|22)p_B(1|2) \nonumber \\
 &&+p(10|12)p_B(0|2) \nonumber \\
 &&+p(00|11)p_B(1|1)+p(01|11)p_B(0|1)+p(01|11)p_B(1|1) \nonumber \\
 &&+p(10|21)p_B(1|1)+p(11|21)p_B(0|1)+p(11|21)p_B(1|1) \nonumber \\
 &&+p(10|21)p_B(0|1) \nonumber \\
 &&+p(00|21)p_B(0|1) \nonumber \\
 &=& \left[p(00|22)+p(01|22)+p(01|22)+p(10|12) \right. \nonumber \\
 &&+p(00|11)+p(01|11)+p(01|11)+p(10|21) \nonumber \\
 && \left.+p(11|21)+p(11|21)+p(10|21)+p(00|21) \right] \frac{1}{2} \nonumber \\
 &=&\frac{6-\sqrt{2}}{4}.
\end{eqnarray}

2.4 With probability $2 \eta^2 (1-\eta)^2$, Alice detects one particle and Bob detects its entangled companion for one of the pairs but none of them detects the particle of the other pair. Then, they output what the detectors in which they found their respective particle. Therefore, the value of $S_2$ for this subensemble is $S_2=\frac{3-\sqrt{2}}{2}$.

2.5 With probability $2 \eta^2 (1-\eta)^2$, Alice detects one particle and Bob detects the one that it is not entangled with it. Therefore, since their outputs are statistically independent, the value of $S_2$ for this subensemble is $S_2=6 \times \frac{1}{4}=\frac{3}{2}$.

2.6 With probability $\eta^2 (1-\eta)^2$, Alice detects the two particles and Bob none (and thus he always outputs~$1$). For this subensemble, $S_2=\frac{1}{4}+0+\frac{1}{4}+\frac{3}{4}+0+0=\frac{5}{4}$.

2.7 With probability $\eta^2 (1-\eta)^2$, Bob detects the two particles and Alice none (and thus she always outputs~$1$). For this subensemble, $S_2=0+\frac{1}{4}+0+\frac{3}{4}+0+\frac{1}{4}=\frac{5}{4}$.

2.8 With probability $2 \eta (1-\eta)^3$, Alice detects one particle and Bob none (and thus, since he observes equal intensities, he outputs~$1$). Therefore, the value of $S_2$ for this subensemble is $S_2=\frac{1}{2}+0+\frac{1}{2}+\frac{1}{2}+0+0=\frac{3}{2}$.

2.9 With probability $2 \eta (1-\eta)^3$, Bob detects one particle and Alice none (and thus, since she observes equal intensities, she outputs~$1$). Therefore, the value of $S_2$ for this subensemble is $S_2=0+\frac{1}{2}+0+\frac{1}{2}+\frac{1}{2}+0=\frac{3}{2}$.

2.10 Finally, with probability, $(1-\eta)^4$, no one detects any particle so each of them outputs~$1$. The value of $S_2$ for this subensemble is $S_2=0+0+0+1+0+0=1$.

Therefore, $\eta_{\text{min}}$ follows from demanding that,
\begin{eqnarray}
&& \eta^4 \left( \frac{21}{16}-\frac{1}{2\sqrt{2}} \right) + 4 \eta^3 (1-\eta) \left(\frac{6-\sqrt{2}}{4}\right) \nonumber \\
&& + 2 \eta^2 (1-\eta)^2 \left(\frac{3-\sqrt{2}}{2}+\frac{5}{4} \right) \nonumber \\
&& + \left[2 \eta^2 (1-\eta)^2+ 4 \eta (1-\eta)^3 \right] \frac{3}{2} \nonumber \\
&& + (1-\eta)^4 < 1,
\end{eqnarray}
which implies
\begin{equation}
\eta_{\text{min}} (N=2) = 0.941.
\end{equation}




Similarly, for the case $N=3$, we found 
\begin{equation}
\eta_{\text{min}} (N=3) = 0.905.
\end{equation}
Calculating $\eta_{\text{min}}(N)$ for higher values of $N$ becomes difficult and it is not really necessary as it is clear that $\eta_{\text{min}}(N)$ will increase with $N$. The reason for this is that the maximum quantum violation rapidly decreases as $N$~increases (see Fig.~\ref{Fig2}), so the fact that the parties yield a quantum-based outcome even when they do not detect all $N$ particles is not enough to account for the possible local hidden variable models.

The problem is that, for Bell experiments with $V \ge 0.98$, the highest experimental detection efficiencies reported are $\eta =0.77$--$0.81$ \cite{SZBS19,LLRZB19}. Therefore, the values for the detection efficiency required to
observe Bell nonlocality based on the intensities produced by $N$ particles are too high for what is achievable with current technology: $\eta \approx 1$ can be achieved \cite{HBD15,RBG17}, but at the cost of visibilities which are not enough for Bell nonlocality based on the intensities of $N>5$ particles.

However, even if the detection efficiency is not enough for a loophole-free Bell test, we can run an experiment adopting the fair sampling assumption. That is, selecting those runs of the experiment in which both parties detect $N$ particles and making the assumption of fair sampling, namely, that the selected runs are a faithful subset of those that would have been obtained if detection efficiency would be perfect. This will allow us to experimentally observe Bell nonlocality using only intensities with current equipment.


\section{Proposed experiment}
\label{exp}


While in scenarios 1 to 3 discussed in Sec.~\ref{mot}, particle pairing information can be, in principle, recovered, there are also scenarios where pairing information is physically erased. Here we introduce a modified Bell test where pairing information is physically erased. Arguably, this is a more convenient experiment for testing the predictions of the previous sections than one in which pairing information exists. The proposed experiment is as follows.

(a) Suppose a source of polarization-entangled pairs of photons based on quantum dots \cite{dot1,dot2} which emits an odd number $N \le 15$ of pairs of entangled photons, with visibility $V>0.98$. The pairs are emitted one by one, with temporal separation~$\tau$ between each pair.

(b) In the space between the source and Alice's measurement device, we introduce beam splitters and mirrors in such a way that photons can go through paths of different lengths. See details later on. In contrast, there is only one possible path between the source and Bob's measurement device.

(c) For each of the local measurements, we use two single-photon detectors, one for each outcome. Each of these detectors must allow us to distinguish two photons that arrive with a time difference $\tau$. If we suitably postselect some runs, we can have no information of which photon of Alice is entangled with which photon of Bob. Notice that, by physically erasing the pairing information before the individual photons are detected one by one, the parties can measure ``changes of intensity'' produced by single photons. 

For example, let us assume for simplicity that $N=3$ and that the three photons of Alice are emitted by the source at times $t_0$, $t_0+\tau$, and $t_0+2 \tau$. Suppose that each of these photons can follow a path of length $6\tau$, or $7\tau$, or $8\tau$, or $9\tau$, or $10\tau$. Now consider those runs in which one photon is detected at $8\tau$, one photon is detected at $9\tau$, and one photon is detected at $10\tau$. In these runs, Alice cannot know which is the photon of Bob each of her photons is entangled with. Still, according to the results in Fig.~\ref{Fig2}, if the visibility is high enough and adopting the assumption of fair-sampling, Alice and Bob can observe a violation of the Bell inequality $S_N \ge 1$ for any $N \le 15$ (with $N$ odd).


\section{Conclusions}


In the ``macroscopic'' limit of infinite number of particles, the violation of Bell inequalities vanishes when the only experimental information are the intensities and we cannot distinguish arbitrarily small differences of intensities. However, here we showed that, for visibilities reachable in current photonic Bell experiments, if the number of photons that reach the detectors every time a local measurement is fixed is $N \le 15$, then Bell nonlocality can be experimentally observed with sufficient statistical significance from the detected intensities, assuming that parties can distinguish any small differences of intensity between their detectors. 

We identified four scenarios in which this result can be useful: Bell experiments based on parametric down-conversion pumped by strong pulses, Bell tests with distant moving sources of entangled pairs and/or local disturbances in the propagation of the particles, Bell experiments using photodetectors based on living cells, and Bell experiments in which particle pairing information is physically erased. In addition, we proposed a Bell experiment of this last type to experimentally test the prediction that Bell nonlocality can be experimentally observed with sufficient statistical significance from the detected intensities whenever $N \le 15$.


\begin{acknowledgments}
We thank B\'arbara Amaral, Jean-Daniel Bancal, Nicolas Brunner, Ana Predojevi\'c, Valerio Scarani, and Giuseppe Vallone for comments. AP was supported by a Grant of the Colegio de Doctorado de F\'{\i}sica del Grupo Tordesillas and \href{http://dx.doi.org/10.13039/501100003593}{CNPq}. AC was supported by \href{http://dx.doi.org/10.13039/100009042}{Universidad de Sevilla} Project Qdisc (Project No.\ US-15097), with FEDER funds, \href{http://dx.doi.org/10.13039/501100001862}{MINECO} Projet No.\ FIS2017-89609-P, with FEDER funds, and QuantERA grant SECRET, by \href{http://dx.doi.org/10.13039/501100001862}{MINECO} (Project No.\ PCI2019-111885-2).
\end{acknowledgments}


\end{document}